\documentclass[11pt,oneside]{article}
\usepackage{a4wide}
\usepackage{mathrsfs}
\usepackage{latexsym,bm}
\usepackage{graphicx}
\usepackage{indentfirst}
\usepackage{slashed}
\usepackage{amsmath}
\usepackage{amssymb}
\usepackage{color}
\usepackage{hyperref}
\usepackage{epsfig}
\usepackage[titletoc]{appendix}
\usepackage{multirow}%
\usepackage{rotating}
\usepackage{cite}
\usepackage{ulem} 
\usepackage[top=2.9cm,bottom=2.5cm,left=2.8cm,right=3cm]{geometry}
\usepackage{tabularx}

\newcommand{\nn}{\nonumber}

\newcommand{\email}[1]{\footnote{{\em } \texttt{#1}}}

\newcommand{\psip}{\psi(3770)}
\newcommand{\xtc}{X(6900)}
\newcommand{\xca}{\chi_{c0}}
\newcommand{\xcb}{\chi_{c1}}
\newcommand{\xcc}{\chi_{c2}}
\newcommand{\hc}{h_c}
\newcommand{\jpsi}{J/\psi}

\begin{document}

\thispagestyle{empty}
\title{
\Large \bf Insights into the inner structures of the fully charmed tetraquark state $X(6900)$ }
\author{\small Zhi-Hui Guo$^{a,b}$\email{zhguo@seu.edu.cn}, \,  J.~A.~Oller$^{c}$\email{oller@um.es}  \\[0.5em]
{ \small\it ${}^a$ School of Physics, Southeast University, Nanjing 211189, China }\\[0.2em]
{ \small\it ${}^b$ Department of Physics and Hebei Advanced Thin Films Laboratory, } \\
{\small\it Hebei Normal University,  Shijiazhuang 050024, China}\\[0.2em] 
{\small {\it ${}^c$ Departamento de F\'{\i}sica. Universidad de Murcia. E-30071 Murcia. Spain}}
}
\date{}

%

\maketitle
\begin{abstract}
  The recently discovered fully charmed tetraquark candidate $X(6900)$ is analyzed within the frameworks of effective-range expansion, compositeness relation and width saturation, and a coupled multichannel dynamical study. By taking into account constraints from heavy-quark spin symmetry, the coupled-channel amplitude including the $J/\psi J/\psi,~ \chi_{c0}\chi_{c0}$ and $\chi_{c1}\chi_{c1}$ is constructed to fit the experimental di-$J/\psi$ event distributions around the energy region near $6.9$~GeV. Another dynamical two-coupled-channel amplitude with the $J/\psi J/\psi$ and $\psi(3770) J/\psi$ is also considered to describe the same datasets. The three different theoretical approaches lead to similar conclusions that the two-meson components do not play dominant roles in the $X(6900)$. Our determinations of the resonance poles in the complex energy plane from the refined coupled-channel study are found to be consistent with the experimental analyses. 
The coupled-channel amplitudes also have another pole corresponding to a narrow resonance $X(6825)$ that we predict sitting below the $\xca\xca$ threshold and of molecular origin.  We give predictions to the line shapes of the $\chi_{c0}\chi_{c0}$ and $\chi_{c1}\chi_{c1}$ channels, which could provide a useful guide for future experimental measurements. 
\end{abstract}

\section{Introduction}

The first fully heavy-flavor tetraquark meson candidate $\xtc$ was recently observed in the di-$\jpsi$ spectra by the LHCb Collaboration~\cite{Aaij:2020fnh}. This intriguing observation has sparked fruitful theoretical  discussions~\cite{x6900refs}. Among the many interpretations, diquark-antidiquark cluster mechanism in the valence-quark picture is currently the most popular theoretical model to explain the narrow state observed around 6.9~GeV~\cite{x6900refs}. Although different theoretical methods, including the quark models, QCD sum rules and effective Lagrangians, could account for the right mass of the tetraquark candidate $\xtc$ without significant difficulty, its $J^{PC}$ quantum numbers and decay patterns are still under vivid debate. Our study aims at pushing forward the clarification of the nature of the $X(6900)$, its possible decay patterns, and the extraction of its resonance pole position from the experimental event distributions by employing a  sophisticated coupled-channel framework based on general principles of $S$-matrix theory. 

In a series of recent works~\cite{Guo:2015daa,Guo:2016wpy,Kang:2016ezb,Kang:2016jxw,Gao:2018jhk,Guo:2019kdc,Meissner:2015mza,Oller:2017alp}, we developed a theoretical framework that is especially useful for bringing insights into the inner structures of the resonance states near some underlying two-hadron thresholds. It is based on the effective-range expansion (ERE) and the compositeness relation \cite{Weinberg:1962hj,Baru:2003qq,Hanhart:2011jz,Hyodo:2011qc,Aceti:2012dd,Sekihara:2014kya,Guo:2015daa,Oller:2017alp} and has been widely and successfully used to study many possible exotic hadrons, such as the charmed baryon $\Lambda_c(2545)$~\cite{Guo:2016wpy}, the hidden bottom tetraquark candidates $Z_b(10610)/Z_b(10650)$~\cite{Kang:2016ezb}, the narrow state $X(3872)$~\cite{Kang:2016jxw}, the pentaquark candidates $P_{c}(4312)$, $P_{c}(4440)$, $P_{c}(4457)$~\cite{Guo:2019kdc} and the hidden charm mesons  $Z_c(3900)$, $X(4020)$, $\chi_{c1}(4140)$, $\psi(4260)$, $\psi(4660)$~\cite{Gao:2018jhk}. This  formalism is especially powerful for describing the elastic single-channel scattering. Just with the inputs of the mass and width of the resonance, we can estimate the scattering length, effective range, and  compositeness coefficient \cite{Guo:2015daa}, whic is the probability of finding the two-hadron component inside the resonance. In this work we will first tentatively apply this formalism to the elastic scattering processes of $\xca\xca$, $\xcb\xcb$ and $\psip\jpsi$ to explore the newly observed $\xtc$. 

Clearly a realistic study needs to include the $\jpsi\jpsi$ channel, in which the $\xtc$ is observed by the experiments~\cite{Aaij:2020fnh}. A similar situation also happens for the pentaquark candidates $P_{c}(4312), P_{c}(4440), P_{c}(4457)$, which requires one to simultaneously include at least the $\jpsi p$ and $\Sigma_c D^{(*)}$ channels~\cite{Guo:2019kdc}. Generally speaking, when the mass of the resonance is below or rather close to the underlying two-hadron threshold, other lighter hadronic degrees of freedom (d.o.f) usually need to be introduced to account for the decay width of the observed resonances. Then a coupled-channel formalism should instead be used in this situation. A simple but efficient approach based on the simultaneous requirements of the decay widths and the compositeness coefficients was developed in Refs.~\cite{Meissner:2015mza,Kang:2016ezb,Guo:2019kdc}. The most important merit of this approach is that we can predict the partial compositeness coefficients (namely, the probabilities of finding the different components inside the considered resonance) and the partial decay widths of the resonance with only the minimal inputs, i.e. the total compositeness of the considered channels, apart from the mass and width of the resonance from the experimental determinations. Within the aforementioned formalism we will perform a three-coupled-channel ($\jpsi\jpsi,\xca\xca, \xcb\xcb$) study and another tentative two-coupled-channel [$\jpsi\jpsi$, $\psip\jpsi$] one  for the $\xtc$. We point out that the quantum numbers $J^{PC}=0^{++}$ are assumed for the $\xtc$ throughout, which is also one of the possibilities for the ground states in the quark model picture~\cite{x6900refs}.

The total compositeness coefficient entering into the above approach should be provided beforehand, and it is usually  taken as a free parameter in practice. To reach more definite conclusions on the nature of the $\xtc$, we further construct the scattering amplitudes involving the $\jpsi\jpsi$, $\xca\xca$ and $\xcb\xcb$ channels by imposing the heavy-quark symmetry to reduce the number of free parameters. We use a general coupled-channel near-threshold (of the $\chi_c\chi_c$ states) parametrization driven by the presence of a Castillejo-Dalitz-Dyson (CDD) pole \cite{castillejo}  without assuming a specific dynamical model, which is indeed more general than the ERE \cite{Kang:2016jxw} or a Flatt\'e parametrization \cite{Kang:2016jxw,han.201031}.  The resulting coupled-channel scattering amplitudes are used to fit the experimental di-$\jpsi$ event distributions by taking into account the rescattering due to the two-hadron systems out of a $\xca\xca$ and $\xcb\xcb$ $S$-wave source.  The resonance pole positions, their couplings to the $\jpsi\jpsi,\xca\xca, \xcb\xcb$ channels and the compositeness coefficients can then be obtained. 
We also give further predictions for the event distributions of the $\xca\xca$ and $\xcb\xcb$.  Subsequently, similar studies involving the $\jpsi\jpsi$ and $\psip\jpsi$ channels are also carried out. 
Our study also predicts the presence of a resonance just below the $\xca\xca$ threshold, named $X(6825)$,  which couples very strongly with the $\xca\xca$ and $\xcb\xcb$ channels and weakly to $\jpsi\jpsi$. As a result, it manifests as a narrow peak at the $\xca\xca$ threshold, being mostly a virtual state of $\xca\xca$ and $\xcb\xcb$.

This paper is organized as follows. First the tentative study of the elastic scattering of $\xca\xca$, $\xcb\xcb$ and $\psip\jpsi$ based on the ERE is explored to address the emergence of the $\xtc$ at its experimental pole position. Then we combine the decay widths and compositeness relations to perform a three-coupled-channel study by assuming several different values for the total compositeness coefficients. Next the fits to the experimental $\jpsi\jpsi$ event distributions with the $\jpsi\jpsi,$ $\xca\xca,$ and $\xcb\xcb$ $S$-wave coupled-channel scattering amplitudes are carried out. As a result, the pole positions of the $\xtc$,   their coupling strengths, total and partial widths, and compositeness coefficients are determined, and compared to those in the previous approaches. We then make the predictions for the line shapes of the $\xca\xca$ and $\xcb\xcb$. Another analysis is also provided by including the coupled $\jpsi\jpsi$ and $\psip\jpsi$ channels, which, however, is disadvantaged relative to the previous case. Finally we give a short summary and conclusions.

\section{Effective range expansion for the elastic scattering }\label{sec.ere}

The standard ERE formalism for the elastic $S$-wave two-body scattering is given by 
\begin{eqnarray}\label{eq.til}
T(E)=\frac{1}{-\frac{1}{a}+\frac{1}{2}r\,k^2-i\,k}\,,
\end{eqnarray}
where $k$ represents the three-momentum in the center of mass (CM) frame, and $a$ and $r$ in this case correspond to the scattering length and the effective range, respectively. The CM three-momentum $k$ in the nonrelativistic limit is related to the CM energy $E$ via 
\begin{eqnarray}
k= \sqrt{2\mu_m(E-m_{\rm th})}\,,
\end{eqnarray}
where the reduced mass $\mu_m$ and the threshold $m_{\rm th}$ are given by 
\begin{eqnarray}
\mu_m=\frac{m_1m_2}{m_1+m_2}\,, \quad  m_{\rm th}=m_1+m_2\,,
\end{eqnarray}
with $m_1$ and $m_2$ the masses of the two scattering particles. It is easy to verify that the ERE amplitude in Eq.~\eqref{eq.til} fulfills the unitarity relation 
\begin{equation}
 {\rm Im}\,T(E)^{-1} = - k  \,,\quad (E > m_{\rm th})\,.
\end{equation}
Loosely speaking, the ERE formula in Eq.~\eqref{eq.til} works well in the energy region sufficiently near the two-particle threshold. 
Three main facts may hinder the applicability of the ERE formalism to the energy region where the resonance lies: the left-hand cuts, the preexisting bare poles \cite{penmor,mor,han.201031,Kang:2016jxw}
(which can also be introduced as CDD poles \cite{castillejo}),   and other important nearby thresholds. Since the exchanges of the color-singlet light hadrons between the two-charmonium states are highly suppressed, the contributions from the left-hand cuts can be safely neglected in the two-charmonium scattering. The situation for the CDD pole is more subtle. If the CDD pole is distant from the two-particle threshold, this is a favorable situation in which to apply the ERE formalism in Eq.~\eqref{eq.til}. While, in the special circumstance in which the CDD pole is close to the threshold, the ERE formula in Eq.~\eqref{eq.til} becomes inaccurate for the description of the near-threshold dynamics~\cite{Guo:2016wpy}. As a consequence, one should explicitly introduce the CDD pole terms into the scattering amplitudes. However, it is usually impracticable to predict
whether or not the CDD poles near the thresholds exist. Alternatively, Refs.~\cite{Guo:2016wpy,Kang:2016ezb} provide an indirect but practical way to discern the validity of the ERE formula. It is obtained in these references 
that when the mass of the CDD pole $M_{CDD}$ approaches the threshold $m_{\rm th}$ the resulting $a$ and $r$ will be linearly and quadratically inversely proportional to $M_{CDD}-m_{\rm th}$ respectively, i.e. 
\begin{eqnarray}
a \propto M_{CDD}-m_{\rm th}\,, \qquad r \propto \frac{1}{(M_{CDD}-m_{\rm th})^2}\,. 
\end{eqnarray}
Since other sources could also contribute to the scattering length,\footnote{The quantum mechanical example of a square well is analyzed in Refs.~\cite{Kang:2016ezb,book.zbs}, where one can observe explicitly that the size of $|a|$ could be very different from the radius $R$ of the square-well potential due to small changes in its depth.} it may not be reliable to infer the CDD pole information from the value of $a$. In contrast, making a comparison with the standard strong interaction scale around 1~fm, a large value of the magnitude for the effective range $r$, would strongly hint at the existence of the near-threshold CDD pole. 
In other words, the large effective range $r$ clearly provides an intuitive and practicable criterion for the existence of the CDD pole near threshold, which also indicates that the ERE formalism in Eq.~\eqref{eq.til} probably is invalid for  describing the dynamics around the threshold. 
On the contrary, if the resulting magnitude of the effective range $r$ is around 1~fm,
it is unlikely that one would necessarily need to introduce  necessarily a CDD pole around the threshold energy region, and the ERE formula in Eq.~\eqref{eq.til} would be sufficient to describe the underlying physics.

Through the analytical continuation one can extrapolate the amplitudes into the second Riemann sheet (RS), where the resonance poles lie. The scattering amplitude in the second RS, $T_{II}(E)$, takes the form   
\begin{eqnarray}\label{eq.tiil}
 T_{II}(E) = \frac{1}{-\frac{1}{a}+\frac{1}{2}r\,k^2\,+i\,k}\,.
\end{eqnarray}
The imaginary part of the three-momentum $k$ should be taken to be positive, i.e. ${\rm Im}k>0$, in Eqs.~\eqref{eq.til} and  \eqref{eq.tiil}. Alternatively, one can still use Eq.~\eqref{eq.til} for $T(E)$, but for calculating $k=\sqrt{2\mu_m(E-m_{\rm th})}$ in the first RS the argument of the radicand is taken between $[0,2\pi)$, while in the second RS it is between $[2\pi,4\pi)$. The resonance pole $E_R = M_R -i\Gamma_R/2$, with $M_R$ the resonance mass and $\Gamma_R$ the width, corresponds to the solution of $T_{II}(E_R)^{-1}=0$, that is, 
\begin{eqnarray}\label{eq.arl}
-\frac{1}{a}+\frac{1}{2}r\,k_R^2+i\,k_R=0\,.
\end{eqnarray}
$k_R$ stands for the three-momentum at the pole $E_R$, i.e. $k_R= \sqrt{2\mu_m(E_R-m_{\rm th})}$. For simplicity, we introduce $k_r$ and $k_i$ to denote the real and imaginary parts of $k_R$, respectively, 
\begin{eqnarray}
k_r={\rm Re}\, k_R\,, \quad  k_i= {\rm Im}\, k_R \,,
\end{eqnarray}
where $k_i>0$ is taken, consistently with the convention for $k$ in Eqs.~\eqref{eq.til} and \eqref{eq.tiil}. It is straightforward to solve Eq.~\eqref{eq.arl} to obtain $a$ and $r$ in terms of $k_r$ and $k_i$. The solutions of $a$ and $r$ were worked out in Ref.~\cite{Guo:2016wpy}, 
\begin{eqnarray}\label{eq.ar0}
a=-\frac{2k_i}{|k_R|^2}\,,\quad r=-\frac{1}{k_i} \,.
\end{eqnarray}
By combining Eqs.~\eqref{eq.ar0} and \eqref{eq.tiil}, the Laurent expansion of the $S$-wave scattering amplitude in the second RS reads~\cite{Kang:2016ezb}  
\begin{eqnarray}
T_{II}(k)=\frac{-k_i/k_r}{k-k_R}+\ldots \,,
\end{eqnarray}
where one can easily identify $-k_i/k_r$ as the residue of the partial-wave amplitude (PWA)  at the pole position in the variable $k$. In our previous study ~\cite{Guo:2015daa,Kang:2016ezb}, the compositeness coefficient $X$, corresponding to the weight of the two-particle component inside the resonance, is shown to be equal to this residue, 
\begin{eqnarray}\label{eq.x}
 X= -\frac{k_i}{k_r}\,.
\end{eqnarray}
It is also proved in Ref.~\cite{Kang:2016ezb} that, when the mass of the resonance pole lies above the considered threshold~\cite{Guo:2015daa}, $X$ in the previous equation is bounded within the range $[0,1]$, and hence it meets the requirement for a probabilistic interpretation.

According to Eqs.~\eqref{eq.ar0} and \eqref{eq.x}, once the mass and width of the resonance are known, the scattering length, effective range, and the compositeness coefficient can be correspondingly predicted under an assumption of single-channel scattering. This assumption in the present case of interest on the $X(6900)$ can be considered simplistic since at least three {\it a priori} relevant channels near the nominal mass of the resonance, the $\xca\xca$, $\xcb\xcb$, and $J/\psi\, \psi(3770)$, can be identified. Their thresholds lie at $6829.4$, $7021.3$ and $6870.6$~MeV, respectively, and have  $\bar{c}\bar{c}cc$ as a valence-quark composition, so that their mutual interactions are Okubo-Zweig-Iizuka (OZI) allowed. 

\begin{table}[htbp]
\centering
\begin{scriptsize}
\begin{tabular}{ c c c c c c c}
\hline\hline
&&&&& \\
Resonance & Mass   & Width & Threshold  & $a$     & $r$  & $X$    \\
           & (MeV)  & (MeV) &  (MeV)    & (fm)    &  (fm)   
\\ \hline \\
$X(6900)$-I & $6905\pm 13$ & $80\pm 38$ & $\xca\xca$~(6829.4)  & $-0.18 \pm 0.07$ & $-1.52 \pm 0.69$ & $0.25 \pm 0.11$  
\\ \hline \\
$X(6900)$-II & $6886\pm 16$ & $168\pm 77$ & $\xca\xca$~(6829.4) & $-0.32 \pm 0.06$ & $-0.72 \pm 0.26$  & $0.53 \pm 0.16$  
\\\hline\hline
\end{tabular}
\end{scriptsize}
\caption{ Scattering length $a$, effective range $r$ and  compositeness coefficient $X$ from the ERE study. The two different sets of mass and width values for the $X(6900)$ are taken from the LHCb determinations~\cite{Aaij:2020fnh}.  \label{tab.ar}} 
\end{table}

We give the numerical results in Table~\ref{tab.ar} for the $\xca\xca$ uncoupled scattering, where the two different sets of masses and widths of the $\xtc$ from the models I and II of LHCb~\cite{Aaij:2020fnh} are separately analyzed. The values of the mass and width of the $X(6900)$ resonance from Ref.~\cite{Aaij:2020fnh} are
\begin{align}
\label{lhcb.mods}
  \text{Model I:}~ M&=6905\pm 11\pm 7~\text{MeV}\,,~ \Gamma=80\pm 19\pm 33~\text{MeV}\,,\\
  \text{Model II:}~M&=6886\pm 11\pm 11~\text{MeV}\,,~\Gamma=168\pm 33\pm 69~\text{MeV}\,,\nonumber
\end{align}
where the distinction is based on the treatment of the nonresonant [with respect to the $X(6900)$] background. For both sets, the masses of the $\xtc$ are below the $\xcb\xcb$ threshold, which is $7021.3$~MeV, and hence we cannot interpret the $X$ defined in Eq.~\eqref{eq.x} as the probability~\cite{Guo:2015daa}. The scattering length and effective range resulting in the elastic $\xcb\xcb$ channel are found to be
\begin{eqnarray}
 a=-0.59\pm0.04\,, \quad  r=-0.31\pm0.02\,,~({\rm case-I})\,,\nonumber \\
a=-0.51\pm0.05\,, \quad  r=-0.28\pm0.02\,,~({\rm case-II})\,,
\end{eqnarray}
 given in units of femtometers.

The small value of the $X$ obtained for $\xca\xca$ scattering indicates that there is an active  extra source of  dynamics beyond the
explicitly included channel, for instance because of the nearby presence of a CDD pole or due to other channels not explicitly included or both
effects simultaneously.
This could hamper the applicability of the elastic ERE to reproduce the pole position of the $X(6900)$, and further study is required. 
On the other hand a value of $X\simeq 1$ would perfectly confirm our onset assumption on the dominance of this channel in order to justify the single-channel treatment.
Therefore, this result clearly indicates that a coupled-channel analysis is required for the $\xtc$.
Similar conclusions can be also made for case II, although the value of $X$ is larger in this case, as it is compatible with the previous one at the level of 1 standard deviation.
The mild values for the effective range $r$ imply that the scenario with a near-threshold CDD pole is disfavored in the uncoupled case.
It should be pointed out that the total width of the $X(6900)$ state is implicitly assumed to be saturated by the $\xca\xca$ or $\xcb\xcb$ in this simplified framework based on assuming the dominance of only one channel within the elastic ERE.
This assumption could be unrealistic, since the partial decay width to the $\jpsi\jpsi$ channel is likely non-negligible.
Therefore, a more realistic study requires the information of the partial decay width to $\xca\xca$, which will be worked out in the next section within a coupled-channel analysis,  after performing fits to the experimental event distributions.

An analogous discussion is in order, when we apply the ERE study to the $\psi(3700)J/\psi$ channel, for which the scattering length $a$, effective range $r$ (given in units of ${\rm fm}$),  and compositeness coefficient $X$ are determined to be 
\begin{eqnarray}\label{eq.arx2ca}
 a=-0.39\pm0.12\,, \quad  r=-1.12\pm 0.46\,, \quad X=0.46\pm 0.18\,,~({\rm case-I})\,,\nonumber \\
a=-0.47\pm0.10\,, \quad  r=-0.57\pm0.16\,, \quad X=0.83\pm 0.17\,,~({\rm case-II})\,.
\end{eqnarray}
The width of the $\psip$ is $27.2\pm 1.0$~MeV \cite{Tanabashi:2018oca} and we take it into account by employing a complex mass, following the approach of Refs.~\cite{Guo:2016wpy,Gao:2018jhk}.  The resulting values found are now 
\begin{eqnarray}\label{eq.arx2cb}
 a=-0.33\pm0.18\,, \quad  r=-1.60\pm1.06\,, \quad X=0.34\pm 0.22\,,~({\rm case-I})\,,\nonumber \\
a=-0.50\pm0.12\,, \quad  r=-0.64\pm0.22\, \quad X=0.80\pm 0.20\,,~({\rm case-II})\,.
\end{eqnarray}
Comparing the two sets of values in Eqs.~\eqref{eq.arx2ca} and \eqref{eq.arx2cb} it is obvious that the finite-width effects from the $\psip$ are small. The resulting values of $a$ and $r$ are similar to those of the $\xca\xca$ case, although the compositeness coefficients become larger because the threshold of the $\psip\jpsi$ is closer to the resonance mass. However, the fact that several important thresholds are involved in this energy region, as indicated above, warns us that the single-channel ERE is probably unreliable, which also agrees with 
the findings in Sec~\ref{sec.threech} as explained therein.

\section{Coupled-channel studies}

In the previous discussion, one notices that the coupled-channel formalism is needed to describe the $\xtc$. We implement two approaches to consider the couplings between several channels, as we detail next.  

\subsection{Three-coupled-channel case}\label{sec.threech}

We consider the $\jpsi\jpsi$ (1), $\xca\xca$ (2) and $\xcb\xcb$ (3) three-coupled channels, with the label for every channel given in parentheses. The contribution from the $\eta_c\eta_c$ channel, whose threshold is rather distant from the energy region of interest around 6.9~GeV, is effectively reabsorbed into the $\jpsi\jpsi$ channel. 
In the infinite quark mass limit, the spin symmetry of the heavy quarks becomes exact, which in turn predicts special patterns for the heavy-hadron spectra. For a heavy-quarkonium system with charm quarks, heavy-quark symmetry predicts that the $\jpsi$ and $\eta_c$ will form a spin doublet, and the $\xca, \xcb, \xcc$, and $\hc$ will form another spin multiplet~\cite{Casalbuoni:1996pg}. These predictions are in reasonable agreement with the experimental measurements~\cite{Tanabashi:2018oca}. To reduce the number of free parameters, we  impose heavy-quark symmetry on the couplings between the charmonia and the $X(6900)$ state. To be more specific, the coupling of the $\jpsi$ pair with $X(6900)$ is denoted as $g_a$. The couplings of the $\chi_{c0}\xca$, $\xcb\xcb$ and the $X(6900)$ are  $g_b$ and $g_b/\sqrt{3}$, respectively, since they are the same aside from a Clebsch-Gordan factor. To properly account for the threshold effects, we take the masses of the charmonia according to their physical values~ \cite{Tanabashi:2018oca}. 

To fix the two couplings $g_a$ and $g_b$, we solve the two equations of the decay width and the compositeness relation. The partial compositeness coefficient $X_j$, i.e. the fraction of the two-particle state of the $j$th channel in the resonance, is given by~\cite{Guo:2015daa}
\begin{eqnarray}\label{eq.xj}
X_j &=& |g_j|^2 \left| \frac{d G_j^{({\rm II})}(s_R)}{d s} \right| \,, 
\end{eqnarray}
where $g_j$ denotes the coupling strength between the two particles in the $j$th channel and the resonance. Depending on whether the pole lies below(above) the $j$th-channel threshold, one should take the two-point one-loop function $G_j(s)$[$G_j^{\rm II}(s)$]  in the first(second) RS. The explicit expression of $G_j(s)$ from dimensional regularization by replacing the divergent term with a subtraction constant takes the form 
\begin{eqnarray}\label{eq.gfunc}
G_j(s)  &=& -\frac{1}{16\pi^2}\left[ a(\mu^2) + \log\frac{m_2^2}{\mu^2}-x_+\log\frac{x_+-1}{x_+}
-x_-\log\frac{x_--1}{x_-} \right]\,, \nonumber\\
 x_\pm &=&\frac{s+m_1^2-m_2^2}{2s}\pm \frac{q_j(s)}{\sqrt{s}}\,.
\end{eqnarray} 
The expression given in Eq.~\eqref{eq.gfunc} represents $G_j(s)$ in the first or physical RS, and its corresponding formula in the second RS reads \cite{npa620}
\begin{eqnarray} 
 G^{\rm II}_j(s) = G_j(s) - i \frac{q_j(s)}{4\pi\sqrt{s}}\,.
\end{eqnarray}
The imaginary part of $G^{\rm II}_j(s)$ has the opposite sign of that of $G_j(s)$ above threshold. 
In Eq~\eqref{eq.gfunc} one has to fix $m_1$ and $m_2$ to the masses of the two particles in this channel. The CM three-momentum in the channel $j$ is $q_j(s)$, given by the standard kinematical expression 
\begin{eqnarray}
q_j(s) =\frac{\sqrt{[s-(m_{j1}+m_{j2})^2][s-(m_{j1}-m_{j2})^2]}}{2\sqrt{s}}\,.
\end{eqnarray}
The function $G_j(s)$ is independent of the regularization scale $\mu$ due to the mutual cancellation of the $\mu$ dependence between the first two terms in Eq.~\eqref{eq.gfunc}, and to be more specific we  set $\mu=770$~MeV throughout this work. Notice that the derivative of the $G_j^{(II)}(s)$ function is independent of the subtraction constant term $a(\mu)$. We note  that there are several proposals in the literature about the extension of  Weinberg's compositeness relation~\cite{Weinberg:1962hj} from the bound state to the resonance situation~\cite{Baru:2003qq,Hanhart:2011jz,Hyodo:2011qc,Aceti:2012dd,Sekihara:2014kya,Guo:2015daa}. We refer to the reader  Ref.~\cite{Gao:2018jhk} for further comparisons between different approaches.

When the mass of the resonance clearly lies above the threshold of the $j$th channel, its partial decay width is calculated by the standard formula~\cite{Tanabashi:2018oca}
\begin{eqnarray} \label{eq.gamma1}
\Gamma_j=  |g_j|^2   \frac{q_j(M_R^2)}{8\pi M_R^2}\,,
\end{eqnarray}
with $q_j(M_R^2)$ being the relativistic CM three momentum  at the resonance mass. In the situation when the mass of the resonance lies close to or even below the $j$th threshold, the above formula is no longer applicable and we introduce the Lorentzian distribution to account for the finite-width effect~\cite{Meissner:2015mza,Kang:2016ezb}, 
\begin{eqnarray}\label{eq.gamma2}
\Gamma_j =  |g_j|^2 \int_{m_{\rm th}}^{M_R+2\Gamma_R} dw \,  \frac{q_j(w^2)}{16\pi^2 \,w^2}  \frac{\Gamma_R}{(M_R-w)^2+\Gamma_R^2/4} \,,
\end{eqnarray}
which naturally recovers the standard decay-width formula~\eqref{eq.gamma1} in the narrow-width limit for $M_R>m_{\rm th}$.  The equations to fix $g_{a}$ and $g_b$ then read
\begin{align}\label{eq.gx}
X= X_1+X_2+X_3 &=|g_a|^2 \left| \frac{d G_1^{\rm II}(s_R)}{d s} \right|+|g_b|^2 \left| \frac{d G_2^{\rm II}(s_R)}{d s} \right|+\frac{|g_b|^2}{3} \left| \frac{d G_3(s_R)}{d s} \right|\,,
\end{align}
and 
\begin{align}\label{eq.gwidth}
\Gamma=\Gamma_1+\Gamma_2+\Gamma_3 &= |g_a|^2  \frac{q_1(M_R^2)}{8\pi M_R^2} +  |g_b|^2 \int_{m_{\rm th, 2}}^{M_R+2\Gamma_R} dw \,\frac{q_2(w^2)}{16\pi^2 \,w^2} \frac{\Gamma_R}{(M_R-w)^2+\Gamma_R^2/4} 
\nonumber \\ &+  \frac{|g_b|^2}{3} \int_{m_{\rm th, 3}}^{M_R+2\Gamma_R} dw \,\frac{q_3(w^2)}{16\pi^2 \,w^2} \frac{\Gamma_R}{(M_R-w)^2+\Gamma_R^2/4} \,,
\end{align}
where the $1/3$ factors in $X_3$ and $\Gamma_3$ correspond to the Clebsch-Gordan coefficient from the angular-momentum superposition.  Equations~\eqref{eq.gx} and \eqref{eq.gwidth} can uniquely determine the coupling strengths $|g_a|$ and $|g_b|$ as a function of the total compositeness $X$. In terms of these couplings we can then calculate the partial decay widths and compositeness coefficients for the  $\jpsi\jpsi$, $\xca\xca$ and $\xcb\xcb$ channels. The results are summarized in Table~\ref{tab.swx1}, where $X$ in Eq.~\eqref{eq.gx} should be taken as an external input (which we also fix below by implementing a coupled-channel dynamical study). It is found that there is a maximum value of $X$, which is 0.4 for the case I and 0.9 for the case II, that allows Eqs.~\eqref{eq.gx} and \eqref{eq.gwidth} to have solutions. We have tried several different values for the $X$.  Larger values of $X$ lead to a smaller magnitude of $|g_a|$ and a bigger one for $|g_b|$. As a result, the partial decay width of the $\jpsi\jpsi$ channel becomes smaller and the decay width of $\xca\xca$ tends to increase. Since the threshold of the $\xcb\xcb$ is clearly higher than the resonance mass, its partial decay width is always very small. With increasing values of $X$, the partial compositeness coefficient of the $\xca\xca$ also becomes larger, while the compositeness values for $\jpsi\jpsi$ and $\xcb\xcb$ always remain small.\footnote{Despite the threshold for $\xcb\xcb$ being clearly larger than $M_R$, we have still calculated the compositeness for this channel because its resulting values are clearly meaningful. This is because they are driven by $|g_b|^2/3$, with $g_b$ being the same as for $\xca\xca$, times the modulus squared of the derivative of the $\xcb\xcb$ $G_3(s)$ function. The latter is also necessarily smaller than for $\xca\xca$ because its threshold is farther  from the resonance pole position.} An analogous analysis by taking the channels $\jpsi\jpsi$ and $\psip\jpsi$ is discussed in Sec.~\ref{sec.210119.1}. 

\begin{table}[htbp]
\centering
\begin{scriptsize}
\begin{tabular}{ c c c c c c c c c c}
\hline\hline
&&&&&&&&\\
Channel & $|g_a|$ & $|g_b|$ & $\Gamma_1$ & $\Gamma_2$ & $\Gamma_3$  & $X_1$   & $X_2$ & $X_3$    \\
          &(GeV)  & (GeV)     & (MeV)      &  (MeV)    &   (MeV)    &    & &
\\ \hline
$X(6900)$-I & & & & & & & & & \\
$X=0.1$   & $7.1_{-2.1}^{+2.2}$ & $6.8_{-0.8}^{+0.6}$ & $64.7_{-33.4}^{+42.5}$  & $15.2_{-4.6}^{+5.3}$ & $0.1_{-0.1}^{+0.3}$ & $0.02_{-0.01}^{+0.02}$ &  $0.06_{-0.01}^{+0.01}$ &  $0.01_{-0.00}^{+0.00}$ \\
$X=0.4$   & $0.6_{-0.5}^{+5.7}$ & $15.4_{-0.5}^{+0.5}$ & $0.4_{-0.4}^{+49.7}$  & $79.2_{-12.5}^{+11.1}$ & $0.4_{-0.4}^{+1.8}$ & $0.00_{-0.01}^{+0.02}$ &  $0.33_{-0.01}^{+0.01}$ &  $0.07_{-0.01}^{+0.01}$ \\
\hline
$X(6900)$-II & & & & & & & & & \\
$X=0.1$   & $11.3_{-3.5}^{+2.7}$ & $5.0_{-3.1}^{+1.6}$ & $160.7_{-83.6}^{+83.3}$  & $7.0_{-6.1}^{+6.8}$ & $0.3_{-0.3}^{+0.1}$ & $0.06_{-0.03}^{+0.03}$ &  $0.03_{-0.03}^{+0.03}$ &  $0.01_{-0.01}^{+0.01}$ \\
$X=0.4$   & $8.9_{-5.5}^{+3.0}$ & $15.0_{-1.0}^{+0.5}$ & $100.9_{-86.2}^{+79.0}$  & $64.3_{-16.9}^{+13.7}$ & $2.8_{-2.6}^{+3.7}$ & $0.04_{-0.03}^{+0.03}$ &  $0.31_{-0.03}^{+0.03}$ &  $0.06_{-0.01}^{+0.01}$ \\
$X=0.9$   & $1.0_{-1.0}^{+6.6}$ & $23.7_{-1.4}^{+1.5}$ & $1.3_{-1.3}^{+71.9}$  & $159.9_{-38.9}^{+44.4}$ & $6.8_{-4.8}^{+10.1}$ & $0.00_{-0.00}^{+0.03}$ &  $0.76_{-0.02}^{+0.02}$ &  $0.14_{-0.02}^{+0.01}$ \\
\hline\hline
\end{tabular}
\end{scriptsize}
\caption{\label{tab.swx1} The solutions of Eqs.~\eqref{eq.gx} and \eqref{eq.gwidth}. The total compositeness coefficient $X$ should be provided as an external input. The maximum values of $X$ for  Eqs.~\eqref{eq.gx} and \eqref{eq.gwidth} admitting solutions are 0.4 for case I and 0.9 for case II, respectively.  } 
\end{table}

To reach a more definite conclusion using the above approach, it is necessary to pin down the value of the total compositeness coefficient $X$, which is, however, not known in advance. 
Another way to proceed is to further constrain the coupling strengths $g_a$ and $g_b$, which will in turn give more definite values for the $X$. In the following, we perform fits to the $\jpsi\jpsi$ invariant-mass distributions from the LHCb Collaboration~\cite{Aaij:2020fnh}, in order to obtain more definite values for the couplings and the resonance pole position. 

Assuming the quantum numbers $J^{PC}=0^{++}$ for the $X(6900)$, the three channels $\jpsi\jpsi$~(1), $\xca\xca$~(2) and $\xcb\xcb$~(3), are all in the $S$ wave. The coupled-channel scattering amplitudes take the form~\cite{Oller:1998zr}
\begin{eqnarray}\label{eq.ut}
 \mathcal{T}(s) = [1-\mathcal{V}(s)\cdot G(s)]^{-1} \cdot \mathcal{V}(s)\,,
\end{eqnarray}
where the right-hand cut is generated through the diagonal matrix $G(s)$, whose diagonal matrix elements are the functions $G_j(s)$ given in Eq.~\eqref{eq.gfunc}. As a result, Eq.~\eqref{eq.ut} satisfies unitarity~ \cite{Oller:1998zr,Oller:2000fj}. The  interacting kernel $\mathcal{V}(s)$ is given by 
\begin{eqnarray} \label{eq.pv}
\mathcal{V}(s)= \left(\begin{matrix}
      0  &  b_{12}  & b_{13} \\
      b_{12} &  \frac{b_{22}}{M_{\jpsi}^2} (s- M_{CDD}^2) & \frac{b_{23}}{M_{\jpsi}^2} (s- M_{CDD}^2)\\
      b_{13} &  \frac{b_{23}}{M_{\jpsi}^2} (s- M_{CDD}^2) & \frac{b_{33}}{M_{\jpsi}^2} (s- M_{CDD}^2)\\
   \end{matrix}\right) \,,
\end{eqnarray}
where the $b_{ij}$ parameters are dimensionless. There is a normalization difference between the unitarized phenomenological amplitude $\mathcal{T}$ in Eq.~\eqref{eq.ut} and the ERE amplitude $T$ in Eq.~\eqref{eq.til}. They are related by $\mathcal{T}=8\pi\sqrt{s}\, T$. To reduce the free parameters, we impose the heavy-quark symmetry to constrain the parameters in the perturbative amplitudes
\begin{eqnarray}
b_{13}=\frac{b_{12}}{\sqrt{3}}\,,\quad b_{23}=\frac{b_{22}}{\sqrt{3}}\,,\quad b_{33}=\frac{b_{22}}{3}\,, 
\end{eqnarray}
where the Clebsch-Gordan coefficients of the angular momenta for $\xca\xca$ and $\xcb\xcb$ have been taken into account. We point out that general parametrizations of the perturbative amplitudes $\mathcal{V}(s)$ have been exploited by introducing polynomial terms, and it turns out that the successful fits generally prefer the form of $s-M_{CDD}^2$ in the $\mathcal{V}(s)$ for the $\xca\xca$ and $\xcb\xcb$ scattering amplitudes, which corresponds to a single CDD pole associated with the unique resonance around it. For the mixing amplitudes $\jpsi\jpsi\to \xca\xca\,,\xcb\xcb$ we find that it is enough to take the constant terms to obtain reasonable fits, as shown in Eq.~\eqref{eq.pv}.  This fact indicates a smooth direct interaction kernel involving the $\jpsi\jpsi$ channel with the other channels, as expected for a far-threshold channel relevant only in providing the total decay width of the resonance.

The formula to describe the experimental $\jpsi\jpsi$ event distributions reads 
\begin{eqnarray}\label{eq.evtdis}
\frac{ d \mathcal{N}(s)} {d \sqrt{s}} = |B_1(s)|^2 \frac{q_{{\jpsi}{\jpsi}}(s)}{M_{\jpsi}^2}\,,
\end{eqnarray}
where the production amplitudes are given by the general parametrization~\cite{Oller:2000fj}
\begin{eqnarray}\label{eq.prodamp}
B(s) = [1-\mathcal{V}(s)\cdot G(s)]^{-1}\cdot \mathcal{P}\,.
\end{eqnarray}
In this equation $\mathcal{P}$ is the vector of production vertexes, which is taken as the constant array
\begin{eqnarray}
 \mathcal{P}= \left(\begin{matrix}
                  d_{1} \\
                  d_{2} \\
                  d_{3} 
                 \end{matrix} \right) \,,
\label{eq.p}
\end{eqnarray}
subsequently modulated by the final-state interactions driven by the $\xca\xca$ and $\xcb\xcb$ strong rescattering \cite{chinogerman,ollerbookrev}. The production parameters $d_j$ are dimensionless, due to the introduction of the $M_{\jpsi}^2$ in Eq.~\eqref{eq.evtdis}. 

Again we impose the heavy-quark symmetry to further constrain the vertexes, so that $d_{3}=d_2/\sqrt{3}$. To be consistent with the assumption in Eq.~\eqref{eq.pv} that the $\jpsi\jpsi$ channel is weakly coupled to the $X(6900)$ state, we set the production vertex $d_1$ to zero and allow $d_2$ to vary only in the fits. It is pointed out that when releasing $d_1$ the fits indeed prefer very large ratios of $d_2/d_1$, but typically give extremely large uncertainties, showing strong correlation between the two parameters. Therefore  our treatment to fix $d_1=0$ not only is motivated by the scattering amplitude~\eqref{eq.pv} but also helps to stabilize the various fits.

For the subtraction constant, its natural value can be estimated by matching the functions $G_j(s)$ calculated in dimensional regularization and with a three-momentum cutoff $q_{\rm max}$  at threshold, as explained in Refs.~\cite{Oller:2000fj,Guo:2018tjx,bco}. This leads to 
\begin{eqnarray}\label{eq.aval}
 a(\mu)=-2\log\left(1+\sqrt{1+\frac{m^2}{q_{\rm max}^2}}\right) + \cdots \simeq -3.0\,,
\end{eqnarray}
by taking $\mu=q_{\rm max}=1.0$~GeV and $m=m_{\xca}$. We will take a universal value for the subtraction constants in the three channels (since masses are rather similar) and fix it to the one given in Eq.~\eqref{eq.aval}. It is further verified that other natural values ranging from $-3$ to $-2$ lead to quite similar results, as explicitly shown later.

Regarding the parameter $M_{CDD}$, we have scanned its values around the range of 6.9~GeV so that the other free parameters are fitted for every value of $M_{CDD}$ fixed, and then there is a clear minimum for the resulting $\chi^2$. To obtain the stable fits, we fix $M_{CDD}$ at the values providing the minimum of $\chi^2$ from the scanning process.

\begin{figure}[htbp]
\centering
\includegraphics[width=0.9\textwidth,angle=-0]{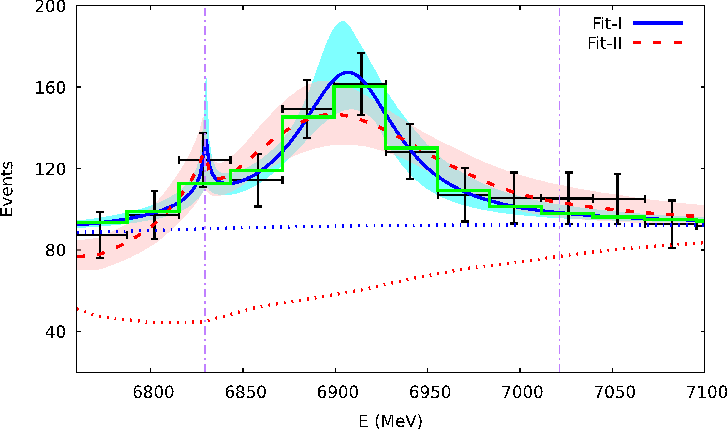} 
\caption{ The $\jpsi\jpsi$ event distribution with the data points taken from Ref.~\cite{Aaij:2020fnh} is shown. The blue and red dotted lines represent the background contributions extracted from models I and II in  Ref.~\cite{Aaij:2020fnh}, respectively. The shaded areas correspond to the error bands at the 1 standard deviation by using the parameters shown in Table~\ref{tab.fitcdd}. 
The histogram given by the green  line  is obtained by averaging  the blue solid line over the bin width for the Fit-I. The left and right vertical lines correspond to the thresholds of $\xca\xca$ and $\xcb\xcb$, respectively. } \label{fig.err1}
\end{figure}

We focus on the experimental data in the energy region around 6.9~GeV~\cite{Aaij:2020fnh}, which amount to 12 data points, as shown in Fig.~\ref{fig.err1}. We take the  background contributions from the experimental analyses~\cite{Aaij:2020fnh}, called there models I and II, to distinguish the two different types of fits that we then perform and which are denoted by Fits I and II, respectively.\footnote{ Here ``background'' denotes all the other contributions in the experimental analysis by the LHCb Collaboration \cite{Aaij:2020fnh} that do not correspond to the resonance signal of the $X(6900)$. The latter is analyzed here within a more sophisticated coupled-channel framework. The main difference between models I and II in Ref.~\cite{Aaij:2020fnh} is that the latter reproduces a dip in the data at around 6.75~GeV.} 
In the fits the free parameters within our approach are finally $b_{12}$, $b_{22}$  in $\mathcal{V}(s)$~\eqref{eq.pv} and $d_2$ in $\mathcal{P}(s)$~\eqref{eq.p}, with $M_{CDD}$ determined as explained.

\begin{table}[htbp]
\centering
\begin{tabular}{ c c c c c c c c c }
\\ \hline\hline
&&&&&& \\
& $\chi^2/{\rm d.o.f}$ & $a(\mu)$ & $M_{CDD}$ & $b_{22}$  & $b_{12}$  & $d_2$   
\\ \hline \\
Fit-I  & $1.6/(12-3)$ & $-3.0^{*}$ & $6910^*$   & $10817_{-2096}^{+8378}$  & $151_{-99}^{+153}$  & $2213_{-316}^{+2106}$  \\  \\
Fit-II & $4.9/(12-3)$ & $-3.0^{*}$ & $6885^*$   & $21085_{-7359}^{+15141}$ & $484_{-112}^{+239}$  & $3646_{-714}^{+1325}$  \\  \\ \hline\hline
\end{tabular}
\caption{\label{tab.fitcdd}   Fit-I and -II obtained with background contributions taken from models I and II in Ref.~\cite{Aaij:2020fnh}, respectively. The entries marked with asterisks are fixed during the fits, as explained in the text, cf. Eq.~\eqref{eq.aval} and the full paragraph following it. $M_{CDD}$ is given in units of MeV, and the parameters $b_{22}$, $b_{12}$, and $d_2$ are dimensionless. } 
\end{table}

We give the outputs of the central fits in Table~\ref{tab.fitcdd}, labeled as Fit-I and Fit-II. For each fit performed $M_{CDD}$ is fixed to some value and, as indicated above, after this scanning there is a clear minimum in the $\chi^2$ for the values  $M_{CDD}=6910$ and 6885~MeV in Fit-I and Fit-II, respectively. With the fitted parameters, we then calculate the resonance pole positions, their residues, and the compositeness coefficients. The resonance poles lie in the complex energy plane of an unphysical RS, which can be accessed via the analytical extrapolation of the $G_j(s)$ functions. Different unphysical RSs of the coupled-channel scattering amplitudes in Eq.~\eqref{eq.ut} can be accessed by properly taking $G_j(s)$ or $G^{\rm II}_j(s)$ for different channels. The second RS is labeled as $(-,+,+)$, where the plus(minus) sign in the $j$th entry  indicates taking $G_j(s)$($G^{\rm II}_j(s))$ in the $j$th channel. In this convention, the first, third,  fourth and fifth sheets are labeled as $(+,+,+)$,  $(-,-,+)$, $(+,-,+)$ and $(-,-,-)$, respectively. The most relevant resonance poles are found to lie in the third RS (which connects continuously with the physical RS between the $\xca\xca$ and $\xcb\xcb$ thresholds). The matrix elements of the  scattering matrix in the unphysical RS around the resonance pole region can be written as 
\begin{eqnarray}
  \label{eq.res}
\mathcal{T}_{kj}(s) = \frac{\gamma_k\gamma_j}{s-M_{\rm pole}^2} + \cdots \,,
\end{eqnarray}
where $\gamma_{k,j=1,2,3}$ are the couplings of the resonance to the corresponding channels and can then be obtained by working out the residue of the PWAs at the resonance pole $M_{\rm pole}$. 
The omitted terms in Eq.~\eqref{eq.res} are the regular parts in the $s-M_{\rm pole}^2$ Laurent expansion. The pole positions and the resonance couplings  $|\gamma_{i=1,2,3}|$ to the different channels are summarized in Table~\ref{tab.pole}. The partial compositeness coefficients $X_{i=1,2,3}$ can be calculated via Eq.~\eqref{eq.xj}, and the results are also given in Table~\ref{tab.pole}. The total compositeness $X$ is given simply by the sum of $X_{i=1,2,3}$. The masses and widths of the resonances from Fit-I and Fit-II are very compatible with the experimental determinations~\cite{Aaij:2020fnh} given in Eq.~\eqref{lhcb.mods}. 
As a result, when taking the masses and widths from the pole positions of the resonances, the ERE parameters in Table~\ref{tab.ar} change only very slightly.

 Let us stress the small value obtained for the total compositeness with $X<0.2$ for the two fits. This fact clearly indicates the dominance of a bare component for the $X(6900)$ and the small weight of the two-hadronic components in its nature. This conclusion  has been reached without assuming any specific dynamical model, instead relying on a general $S$-matrix parametrization, Eqs.~\eqref{eq.ut} and \eqref{eq.pv}.  The smallness of $X$ is a reflection of the  value of $M_{CDD}$ lying so close to  the resonance mass \cite{Kang:2016jxw,mor}. This is the basic point stressed in the Morgan's counting-pole criterion on the nature of a resonance \cite{mor,penmor}, because it drives to the proliferation of similar pole positions in  different RSs \cite{mor}.  We have checked that this is the case here too, and poles  are found in different RSs associated with the inelastic channel $\xcb\xcb$ with little variation in their positions, as required by this criterion.

\begin{table}[htbp]
\centering
\begin{scriptsize}
\begin{tabular}{ c c c c c c c c c c c }
\\ \hline\hline
&&&&&&&&&\\
& Mass & Width/2 & $|\gamma_1|$ & $|\gamma_2|$ & $|\gamma_3|$ & $X_1$  & $X_2$  & $X_3$  &  $X=\sum_{i=1}^{3} X_i$   
\\     &(MeV)  & (MeV)     & (GeV)      &  (GeV)    &   (GeV)    &    & & 
\\  &   &       &      &     &      &    & &  \\
 Fit-I   & $6907_{-3}^{+5}$ & $33_{-10}^{+14}$ & $4.6_{-2.8}^{+2.5}$  & $9.7_{-2.6}^{+1.4}$ & $5.6_{-1.5}^{+0.8}$  & $0.01_{-0.01}^{+0.01}$ & $0.13_{-0.06}^{+0.04}$ & $0.03_{-0.01}^{+0.01}$ & $0.17_{-0.07}^{+0.04}$  \\
   &   &       &      &     &      &    & &  \\
 Fit-II  & $6892_{-2}^{+2}$ & $80_{-17}^{+24}$ & $10.3_{-1.4}^{+1.8}$  & $6.9_{-1.9}^{+1.4}$ & $4.0_{-1.1}^{+0.8}$  & $0.05_{-0.01}^{+0.02}$ & $0.06_{-0.03}^{+0.03}$ & $0.01_{-0.01}^{+0.01}$ & $0.13_{-0.03}^{+0.03}$ 
\\ \hline\hline
\end{tabular}
\end{scriptsize}
\caption{  The resonance poles of the $X(6900)$ in the third Riemann sheet from the three-coupled-channel fits are given. $|\gamma_{i=1,2,3}|$ represent the resonance couplings to the $\jpsi\jpsi, \xca\xca$ and $\xcb\xcb$ channels, respectively. $X_{i=1,2,3}$ denote the partial compositeness coefficients, i.e., the probabilities of finding the $\jpsi\jpsi$, $\xca\xca$ and $\xcb\xcb$ components in the $X(6900)$ state, respectively. 
\label{tab.pole} } 
\end{table}


The resulting fits are plotted in Fig.~\ref{fig.err1}, where the error bands at the 1 standard deviation for the Fit-I (blue solid line) and Fit-II (red dashed line), are provided.  The blue and red dotted lines in Fig.~\ref{fig.err1} represent the background contributions of models I and II, respectively, extracted from Ref.~\cite{Aaij:2020fnh}.
A cusp effect is clearly seen at the threshold of the $\xca\xca$ indicated by the vertical left line, while the right one corresponds to the $\xcb\xcb$ threshold. 
However, this narrow peak washes out when performing the averages over the width  around 27~MeV of the experimental energy bins.
This is explicitly shown by the histogram in Fig.~\ref{fig.err1}, which results were obtained by performing the average inside each bin width.
So as not to overload the plot, only the histogram obtained in this way for Fit-I is explicitly shown. 
The histogram from Fit-II shows quite a similar trend. 

Indeed, we have checked to see that this remarkable  cusp effect in the $\jpsi\jpsi$ event distribution obtained with our results is unveiling the presence of a pole quite close to the $\xca\xca$ threshold which enhances its contributions. Of course, future experimental measurements with higher statistics and better energy resolution in the narrow energy region around the $\chi_{c0}\chi_{c0}$ threshold will then be crucial in order to definitely  discriminate this possible cusp effect and associated resonance. The existence of this new fully charmed tetraquark resonance, which we name $X(6825)$, is a neat prediction of our dynamical approach that should be considered in future experimental searches.

The pole of the $X(6825)$ lies in fourth RS, $(+,-,+)$, and its positions and residues obtained from the different fits are (primes are affixed to all the symbols referring to this resonance):\footnote{The fourth RS $(+,-,+)$ connects continuously with the second RS $(-,+,+)$ by crossing the physical $s$ axis above the $\xca\xca$ threshold. In turn the second RS connects continuously with the physical one below the same threshold.}
\begin{align}
&\text{Values for  Fit-I:} \nn\\
\label{eq.poleiv1}
& E'_R=6827.0_{-4.8}^{+1.6}-i1.1_{-1.0}^{+1.3}\,, \,\,\, |\gamma'_1|=1.4_{-0.9}^{+0.6}\,,\,\,\, |\gamma'_2|=11.9_{-3.1}^{+3.2}\,, \,\,\, |\gamma'_3|=6.8_{-1.8}^{+1.8}\,,\\
&\text{Values for Fit-II:} \nn\\
&E'_R=6820.6_{-2.7}^{+3.0}-i4.0_{-1.6}^{+1.7}\,, \,\,\, |\gamma'_1|=2.5_{-0.6}^{+0.5}\,,\,\,\, |\gamma'_2|=15.8_{-0.6}^{+0.7}\,, \,\,\, |\gamma'_3|=9.1_{-0.4}^{+0.4}\,,\nn
\end{align}
where $E'_R$ is given in MeV and the residues are given in GeV. 
Comparing them to the residues of the $X(6900)$ in Table~\ref{tab.pole}, we see that the $X(6825)$ couples even more strongly to the $\xca\xca$ and $\xcb\xcb$ channels. The couplings to these channels follow the rule that $|\gamma'_3|\approx |\gamma'_2|/\sqrt{3}$, according to the heavy-quark symmetry. However, the coupling to $\jpsi\jpsi$ is much weaker and this fact explains the much smaller width of the $X(6825)$, identified as minus twice the imaginary part of its pole position, as compared with the $X(6900)$. One can semiquantitatively understand the magnitude of the ratio $|\gamma_1/\gamma'_1|$, by assuming dominance of the $\jpsi\jpsi$ decay width for the $X(6825)$ and $X(6900)$. In this way, according to Eq.~\eqref{eq.gamma1} one has
\begin{align}
\label{210123.1}
\left|\frac{\gamma_1}{\gamma'_1}\right|\approx \sqrt{\frac{ q(M'_R)\Gamma}{q(M_R)\Gamma'}}~, 
\end{align}
whose central value is equal to 5.3 and 4.3 for Fit-I and -II, respectively. These figures are pretty much compatible with the numbers of $|\gamma_1/\gamma'_1|$ given in Table~\ref{tab.pole} and Eq.~\eqref{eq.poleiv1}.
The facts that the pole couples  strongly to the $\xca\xca$ and $\xcb\xcb$ channels and that it lies so close to the $\xca\xca$ threshold in the $(+,-,+)$ sheet, are clear indications that the $X(6825)$ is a virtual state made up predominantly of these channels (loosely speaking, this is usually referred as ``molecular origin''). Regarding the application of Morgan's counting-pole criterion, the pole position of this resonance lies only in the sheet indicated, which is also in agreement with this interpretation. Indeed, in the limit of $b_{12}\to 0$, while keeping the values for $b_{22}$, $M_{CDD}$ and $a(\mu)$ from the fits, the coupling to $\jpsi\jpsi$ is zero and the pole becomes a pure virtual state, around 6827 and 6825~MeV for Fit-I and Fit-II respectively. Another interesting limit is to take the mass of the $\xcb$ equal to the physical mass of the $\xca$. Then the pole moves to the first RS and it is a bound state. From this point if $b_{12}$ is taken back to its fit value in Table~\ref{tab.fitcdd} then this bound state becomes a resonance in the second RS due to the coupling to $\jpsi\jpsi$ with finite width. By continuously increasing the mass of the $\xcb$, the jump of the pole to the fourth RS is seen slightly above the $\xca\xca$ threshold and then, after a small increase in $m_{\xcb}$, the pole turns left and finally evolves up to its final position in Eq.~\eqref{eq.poleiv1}, below the two $\xca$ threshold for the physical mass of the $\xcb$. 

\begin{figure}[htbp]
\centering
\includegraphics[width=0.95\textwidth,angle=-0]{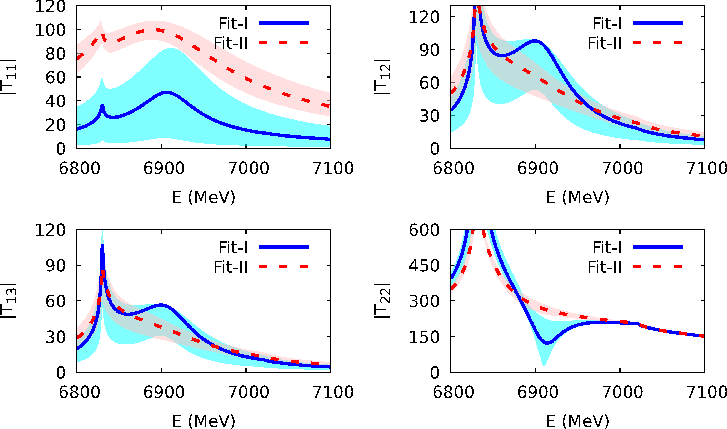} 
\caption{ Our predictions for the scattering amplitudes. The top left and right panels show the amplitudes of $\jpsi\jpsi\to\jpsi\jpsi$ and $\jpsi\jpsi\to\xca\xca$. The bottom left and right panels give the predictions for the amplitudes of $\jpsi\jpsi\to\xcb\xcb$ and  $\xca\xca\to\xca\xca$, respectively. Since the heavy-quark symmetry is imposed, the shapes of the amplitudes involving $\xcb$ look rather similar to those with $\xca$ and we do not explicitly show other amplitudes. }
   \label{fig.err2}
\end{figure}

Due to the large couplings of the $X(6825)$ to $\xca\xca$ and $\xcb\xcb$ one would expect a prominent peak at the $\xca\xca$ threshold in the PWAs involving these channels. However, its form is asymmetric because of the $\xca\xca$ threshold.
The predictions for the scattering amplitudes from different fits are shown explicitly in Fig.~\ref{fig.err2},
where the shaded areas correspond to the error bands at the 1$\sigma$ level from the Fit-I and Fit-II in Table~\ref{tab.fitcdd}.
Strong cusp effects around the $\xca\xca$ threshold are clearly seen in all the scattering amplitudes, though, as expected, this peak is much less prominent for $T_{11}$. It is much more relevant for the $T_{12}$ and $T_{13}$, and  huge in the diagonal $T_{22}$ and $T_{33}$. Of course,
this hierarchy is not  more than a reflection of the sizes of the $|\gamma'_i|$ given in Eq.~\eqref{210123.1}. It is also visible in  Fig.~\ref{fig.err2} that the peak at the cusp is broader for Fit-II than for Fit-I.  
Regarding the $X(6900)$ in Fig.~\ref{fig.err2}, for  Fit-I the resonant enhancement obviously shows up in the amplitudes $\jpsi\jpsi\to \jpsi\jpsi, \xca\xca$ and $\xcb\xcb$, 
while in the transition amplitudes $\chi_{0,1}\chi_{0,1}\to\chi_{0,1}\chi_{0,1}$ the resonance manifests as a dip due to the destructive interference with the $X(6825)$. 
For the Fit-II case, except for $\jpsi\jpsi\to\jpsi\jpsi$, the $X(6900)$   barely shows any structure in the scattering amplitudes, which seems to be  consistent with the rather large widths from Fit-II [we can say that the wide $X(6900)$ for Fit-II is eaten up by the $X(6825)$ strong signal  in the amplitudes involving the channels $\xca\xca$ and $\xcb\xcb$].

Other fits by taking different values of the subtraction constants $a(\mu)$ are also obtained. To make a clear comparison of the fits in Table~\ref{tab.fitcdd}, the values of $M_{CDD}$ of the Fit-I and Fit-II will be fixed at the same values as shown in the previous table. For the Fit-IA by fixing $a(\mu)=-2.5$, the fitted parameters, $\chi^2$, resonance pole position $E_R$ and $X$ are (in order): $b_{22}=5234.6$, $b_{12}=155.0$ and $d_2=1229.7$,  $\chi^2=3.5$, $E_R=6925-i\,41$~MeV and $X=0.11$.
In the same way for Fit-IIA we have:  $b_{22}=3442.7$, $b_{12}=200.0$ and $d_2=1549.6$,  $\chi^2=7.7$, $E_R=6909-i\,99$~MeV and  $X=0.23$.
We now take $a(\mu)=-3.5$ which is below the nominal value $-3.$ Fit-IB and its results are $b_{22}=45382.3$,  $b_{12}=538.0$ and $d_2=3353.1$, $\chi^2=3.5$, $E_R=6923-i\,42$~MeV and  $X=0.04$.  For Fit-IIB one has  $b_{22}=46782.2$, $b_{12}=871.3$ and $d_2=5734.4$, $\chi^2=10.1$,  $E_R=6879.5-i\,104$~MeV and $X=0.09$. Here we show only the most relevant resonance poles found in the third RS for the $X(6900)$. It is noticeable that in all cases the total compositeness coefficients $X$ clearly remain  small, the pole positions vary only moderately  and the $\chi^2$ gets worse by around a factor of 2.
The resulting curves obtained in the scenarios of the Fit-IA/B and Fit-IIA/B are not explicitly shown, since they look quite similar as those from the Fit-I and Fit-II, respectively. In this way we verify that  different fits by taking different values for the subtraction constants $a(\mu)$ within the same type of background contributions lead to rather similar results for the di-$\jpsi$ event distributions, while the two different types of fits by using different models for the background, as provided by the LHCb analyses \cite{Aaij:2020fnh},  give obviously different results.

Since now the total compositeness coefficient $X$ is known from the fits, it is interesting to redo the analyses by combining Eqs.~\eqref{eq.gx} and \eqref{eq.gwidth}. For case I when fixing $X=0.17$, the solutions of Eqs.~\eqref{eq.gx} and \eqref{eq.gwidth} read 
\begin{eqnarray}\label{eq.gxwresli}
&& |g_a|=6.2~{\rm GeV}\,, \quad |g_b|=9.5~{\rm GeV}\,, \quad \Gamma_1=49.7~{\rm MeV}\,, \quad\Gamma_2=30.1~{\rm MeV}\,, \quad\Gamma_3=0.2~{\rm MeV}\,, \nonumber \\
&& X_1=0.018\,, \quad X_2=0.126\,, \quad X_3=0.026\,,
\end{eqnarray}
which agree well with the Fit-I results from the sophisticated coupled-channel study in Table~\ref{tab.pole}. For case II when fixing $X=0.13$, the solutions of Eqs.~\eqref{eq.gx} and \eqref{eq.gwidth} are 
\begin{eqnarray}\label{eq.gxwreslii}
&& |g_a|=11.1~{\rm GeV}\,, \quad |g_b|=6.7~{\rm GeV}\,, \quad \Gamma_1=154.7~{\rm MeV}\,, \quad\Gamma_2=12.8~{\rm MeV}\,, \quad\Gamma_3=0.5~{\rm MeV}\,, \nonumber \\
&& X_1=0.06\,, \quad X_2=0.06\,, \quad X_3=0.01\,,
\end{eqnarray}
which are also in good accord with the Fit-II results in Table~\ref{tab.pole}. When obtaining the values in Eqs.~\eqref{eq.gxwresli} and \eqref{eq.gxwreslii}, we take the masses and widths of the $X(6900)$ from the experimental analyses in Ref.~\cite{Aaij:2020fnh}. These equations also provide the partial decay widths of the $X(6900)$ to the different channels. 
It is worth pointing out that when using the masses and widths of the resonances from our fits in Table~\ref{tab.pole}, the agreement of the residues and partial compositeness coefficients turn out to be excellent. Therefore, we provide here a solid demonstration that the coupled-channel methods by utilizing the compositeness relations and the decay widths, namely, Eqs.~\eqref{eq.gx} and \eqref{eq.gwidth}, indeed offer a very convenient and reliable approach to studying the resonance dynamics. 


When we consider the results in Table~\ref{tab.ar} from the single-channel ERE study, it seems that the compositeness coefficients predicted by the elastic ERE are larger and in contradiction to the results in Table~\ref{tab.pole} from the coupled-channel fits. However, it should be stressed that in the elastic ERE study the total width of the $X(6900)$ is assumed to be saturated by the $\xca\xca$ channel, which does not seem to be consistent with the partial decay width predicted by the coupling/residue in Table~\ref{tab.pole}. To be more specific, we calculate next the partial decay width to $\xca\xca$, $\Gamma_2$, by removing from the central values of the total width in Table~\ref{tab.pole} the  easily calculable  $\jpsi\jpsi$ decay widths in terms of the central values of $|\gamma_1|$, cf. Eq.~\eqref{eq.gamma1} with $|g_1|$ substituted for $|\gamma_1|$. The relative uncertainty for the resulting $\Gamma_2$ is estimated as twice the relative error for $|\gamma_2|$ (because it depends quadratically on the coupling). We then obtain the values 
\begin{eqnarray}\label{eq.newwidth}
\Gamma_{2}=40^{+11}_{-20} ~{\rm MeV}\,\,({\rm Fit-I})\,,\quad \Gamma_{2}=26^{+11}_{-14}~{\rm MeV}\,\,({\rm Fit-II})\,.
\end{eqnarray}
By taking these realistic partial decay widths, the central values for the single-channel ERE parameters turn out to be 
\begin{eqnarray}
 a=-0.10~{\rm fm}\,, \quad r=-3.0~{\rm fm}\,, \quad X=0.13 \,\, (\rm Fit-I)\,,
\end{eqnarray}
and 
\begin{eqnarray}
 a=-0.09~{\rm fm}\,, \quad r=-4.1~{\rm fm}\,, \quad X=0.10 \,\,(\rm Fit-II)\,.
\end{eqnarray}
It is clear that the compositeness coefficients from the ERE are now much smaller than in Table~\ref{tab.ar}, as they should be, and indeed they are fairly compatible with the results for $X_2$ in Table~\ref{tab.pole}  from the fits.

\begin{figure}[htbp]
\centering
\includegraphics[width=0.95\textwidth,angle=-0]{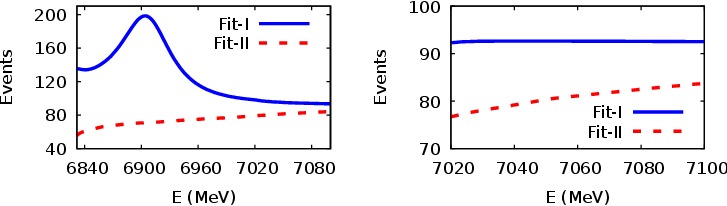} 
\caption{  Our predictions for the distributions of (left panel) $\xca\xca$ and (right panel) $\xcb\xcb$.  }
   \label{fig.err1b}
\end{figure}

Therefore, after taking a more sophisticated coupled-channel analysis, our results confirm the findings of Ref.~\cite{Aaij:2020fnh} for the $X(6900)$. The resonance pole positions from Fit-I and Fit-II are very compatible with the masses and widths determined in models I and II in Ref.~\cite{Aaij:2020fnh}, respectively. As a novelty, our coupled-channel study provides new information on the couplings of the $\xca\xca$ and $\xcb\xcb$, which in turn allows us to calculate the partial compositeness coefficients; cf.  Eqs.~\eqref{eq.gxwresli} and \eqref{eq.gxwreslii}. Furthermore, the coupled-channel analyses also enable us to predict the line shapes of the distributions of the $\xca\xca$ and $\xcb\xcb$, as shown in Fig.~\ref{fig.err1b}, which could provide useful guidelines for the experimental study in the next step.  Future measurements on the distributions of the $\xca\xca$ and $\xcb\xcb$ will definitely be helpful for discriminating among different scenarios proposed here and for reaching a more definite conclusion for the properties of the $X(6900)$ state, since the $\xca\xca$ and $\xcb\xcb$ event distributions look very different for Fit-I and Fit-II, particularly for the former channel.

\subsection{Coupled-channel systems with $\psip\jpsi$}
\label{sec.210119.1}

Other channels that one could include in the coupled-channel study are the  $\psi(2S)\jpsi$ and $\psip\jpsi$~\cite{x6900refs}, since their mutual interactions and those with the channels $\jpsi\jpsi$,  $\xca\xca$ and $\xcb\xcb$ are OZI allowed.
As a complementary study to the discussions in the previous section, coupled-channel analyses including the $\psip\jpsi$ channel (whose threshold is 6870.6~MeV \cite{Tanabashi:2018oca}) will be further explored.
We cannot afford to study simultaneously all the four channels $\jpsi\jpsi$, $\xca\xca$, $\xcb\xcb$ and $\psip\jpsi$ due to the proliferation of free parameters.\footnote{At least four more free parameters would be needed when  following a similar scheme as before, namely, $b_{12}$, $b_{14}$, $b_{22}$, $b_{24}$, $b_{44}$, $d_2$ and $d_4$, with the subscript 4 referring to the $\psip\jpsi$ channel.}. 
Because of this, and also due to the fact that the threshold of  $\psi(2S)\jpsi$  is at 6783.0~MeV \cite{Tanabashi:2018oca}  below the energy region for the resonance signal at the lower end of Fig.~\ref{fig.err1}, we do not  further consider this  channel.

We again consider the quantum numbers $J^{PC}=0^{++}$ for the $X(6900)$ and study the $S$-wave scattering with the two coupled channels  $\jpsi\jpsi$~(1) and $\psip\jpsi$~(2). We follow the same formalism as in the previous section and write the perturbative kernel as 
\begin{eqnarray} \label{eq.pv2}
\hat{\mathcal{V}}(s)= \left(\begin{matrix}
      0  &  \hat{b}_{12}  \\
      \hat{b}_{12} &  \frac{\hat{b}_{22}}{M_{\jpsi}^2} (s- \hat{M}_{CDD}^2) 
   \end{matrix}\right) \,,
\end{eqnarray}
where the hat symbol is introduced over the parameters in order to distinguish them from those in Eq.~\eqref{eq.pv} that are used for the  three-coupled-channel case. We use this same convention to distinguish everything between these two scenarios.   
Needless to say, the unitarized scattering amplitudes and the production amplitudes will be  constructed according to Eqs.~\eqref{eq.ut} and \eqref{eq.prodamp}, respectively, with the obvious changes of the corresponding thresholds and relevant parameters. The subtraction constant $\hat{a}(\mu)$ is fixed by taking $q_{\rm max}=1$~GeV and $m=m_{\psip}$, which leads to $\hat{a}(\mu)=-3.2$.
Taking exactly the same background terms and fit strategies as the previous three-channel study, we give the parameters from the two-channel fits in Table~\ref{tab.fitcdd2c}.
This fit is not well fixed, as one can clearly infer because of the large uncertainties in the free parameters $\hat{b}_{12}$, $\hat{b}_{22}$, and $\hat{d}_2$ written in Table~\ref{tab.fitcdd2c}, which are essentially undetermined due to their large mutual correlations. Furthermore, the fit is also unstable around the threshold of the $\psip\jpsi$, giving rise to quite different $\jpsi\jpsi$ event distributions in this energy region that somewhat wash out when averaging along the width of the experimental bin. This fact potentially disfavors this type of fits in relation to the three-coupled-channel ones.

Nonetheless, we give for reference the resulting resonance poles, residues, and the compositeness coefficients for the two-coupled-channel fits in Table~\ref{tab.pole2c} so that some trends are common to all the explored fits of this kind. 
One can observe that the CDD locations for Fit-I and -II are similar in  Tables~\ref{tab.pole} and \ref{tab.pole2c}, and that they lie almost on top of the resonance mass. Indeed, the pole positions given are also close to each other, with the biggest difference being  the discrepancy between the masses of the resonance in the Fit-II case, although it is much smaller than the rather large width of the resonance, so this  fact is not really relevant. 
It is also the case in the two-coupled channel analysis that virtual poles are present close to the $\psip\jpsi$ threshold at around 6855 for Fit-I and Fit-II, similarly to the $X(6825)$ in the three-coupled-channel case. 
The conclusion that the $X(6900)$ resonance originated mainly from the bare CDD pole in our study is quite robust, regardless of the dynamical channels included near the energy region around 6.9~GeV. As a matter of fact, the value of $\hat{X}$ in the two-coupled-channel study  is  even smaller than in the three-coupled-channel case.
For the two-coupled-channel case $\hat{X}_1$ is the largest, with a much smaller $\hat{X}_2$,  while for the three-coupled-channel scattering the opposite situation occurs.

Another interesting fact that points toward a preference for the the three-coupled-channel case over the two-coupled-channel one is the magnitude of the couplings $|\gamma_2|$ and $|\hat{\gamma_2}|$. If one compares their central values given in Tables~\ref{tab.pole} and \ref{tab.pole2c}, the result is that the  values for $|\gamma_2|$ are around a factor of 5 (Fit-I) and 3.5 (Fit-II) larger than for  $|\hat{\gamma}_2|$. This is quite a difference and it indicates that the $X(6900)$ seems to couple much more strongly to $\xca\xca$, and also to $\xcb\xcb$, than to $\psip\jpsi$. This result would justify considering the reduction of possible channels more realistic in the three-coupled-channel scenario than to the two-coupled-channel one. The smallness of the residue $|\hat{\gamma}_2|$ is caused because of the large value of the parameter $\hat{b}_{22}$, which tends to be more than an order of magnitude larger than $b_{22}$. 

It is also interesting to redo the analyses of the saturation of width and compositeness in Eqs.~\eqref{eq.gx} and \eqref{eq.gwidth} in terms of the channels $\jpsi\jpsi$ and $\psip\jpsi$.
By imposing the total compositeness coefficients from the fits in Table~\ref{tab.fitcdd2c}, it is straightforward to solve the equations that follow from the saturation of the width and compositeness. The solutions are 
\begin{align}\label{eq.gxwreslic2}
&\text{$\hat{X}$ from Fit-IA:}\nonumber\\
  |\hat{g}_1|&=7.9~{\rm GeV}, \,\, |\hat{g}_2|=2.5~{\rm GeV}, \,\, \hat{\Gamma}_1=78.6~{\rm MeV}, \,\, \hat{\Gamma}_2=1.4~{\rm MeV}, \,\,   \hat{X}_1=0.03, \,\, \hat{X}_2=0.01\,,\\ 
&\text{$\hat{X}$ from Fit-IIA:}\nonumber\\
|\hat{g}_1|&=11.5~{\rm GeV}, \,\, |\hat{g}_2|=2.2~{\rm GeV}, \,\, \hat{\Gamma}_1=167.0~{\rm MeV}, \,\, \hat{\Gamma}_2=1.0~{\rm MeV}, \,\, \hat{X}_1=0.06, \,\, \hat{X}_2=0.01\,,
\end{align}
which are consistent with the results for Fit-$\hat{\rm I}$ and Fit-$\hat{\rm II}$ given in Table~\ref{tab.pole2c}.
To obtain these values  
the experimental masses and widths of the $X(6900)$ from Ref.~\cite{Aaij:2020fnh} are used. 
We further verify that the results are barely affected when the masses and widths are taken from the resonance poles given in Table~\ref{tab.pole2c}.

As mentioned above a more sophisticated coupled-channel analysis would be to simultaneously include the four dynamical channels $\jpsi\jpsi$~(1), $\xca\xca$~(2), $\xcb\xcb$~(3) and $\psip\jpsi$~(4).
The biggest challenge in this case is the large number of free parameters. 
To perform a check of stability of our previous results with the three-coupled channels,  
we treat the $\psip\jpsi$ channel in a perturbative manner similar to the $\jpsi\jpsi$, so that $b_{14}=b_{44}=0$ 
and $b_{12}$, $b_{24}$, and $d_{2}$ are allowed to float, with
$b_{34}=b_{24}/\sqrt{3}$, $b_{13}=b_{12}/\sqrt{3}$ and $d_3=d_2/\sqrt{3}$, because of the heavy-quark symmetry. 
In this way, only one additional free parameter is needed, see the discussions in Sec.~\ref{sec.threech}. 
The resulting parameters and the results from such four-channel fits turn out to be almost identical to the values in Tables~\ref{tab.fitcdd} and \ref{tab.pole}, respectively, which points favorably toward the stability of the results from the three-coupled-channel study.

 Alternatively, we could switch the roles of the $\xca\xca, \xcb\xcb$ channels and the $\psip\jpsi$ one and include the two former channels perturbatively in the two-coupled-channel study. However, the fit does not become more stable than  the previous one with only the $\jpsi\jpsi$ and $\psip\jpsi$ channels, and the fit results are essentially the same as in Table~\ref{tab.pole2c}  (despite having one more free parameter). Therefore, we do not dwell further on these results and refer to the discussion above.

\begin{table}[htbp]
\centering
\begin{tabular}{ c c c c c c c c c }
\\ \hline\hline
&&&&&&\\
& $\chi^2/{\rm d.o.f}$ & $\hat{a}(\mu)$ & $\hat{M}_{CDD}$ & $\hat{b}_{22}$  & $\hat{b}_{12}$  & $\hat{d}_2$ 
\\ \hline \\
Fit-$\hat{\rm I}$  & $2.8/(12-3)$ & $-3.2^{*}$ & $6900^*$   & ${(2.4_{-1.7}^{+4.6})}\cdot10^{5}$  & $1303_{-597}^{+1243}$  & $7825_{-3495}^{+6318}$  \\  \\
Fit-$\hat{\rm II}$ & $2.4/(12-3)$ & $-3.2^{*}$ & $6880^*$   & ${(1.5_{-0.6}^{+1.8})}\cdot10^{5}$ & $1356_{-305}^{+741}$  & $9675_{-2043}^{+4674}$ \\   \hline\hline
\end{tabular}
\caption{\label{tab.fitcdd2c} Results from the two-channel fits including the $\jpsi\jpsi$ and $\psip\jpsi$. } 
\end{table}

\begin{table}[htbp]
\centering
\begin{tabular}{ c c c c c c c c c }
\\ \hline\hline
&&&&&&&\\
& Mass & Width/2 & $|\hat{\gamma}_1|$ & $|\hat{\gamma}_2|$   & $\hat{X}_1$  & $\hat{X}_2$  &  $\hat{X}=\sum_{i=1}^{2} \hat{X}_i$   
\\     &(MeV)  & (MeV)     & (GeV)      &  (GeV)    &      & & 
\\ \hline \\
 Fit-$\hat{\rm I}$   & $6900_{-1}^{+2}$ & $44_{-16}^{+20}$  & $8.2_{-1.6}^{+1.7}$ & $1.9_{-0.9}^{+1.5}$  & $0.03_{-0.01}^{+0.01}$ & $0.01_{-0.00}^{+0.01}$  & $0.04_{-0.01}^{+0.02}$  \\
   &   &       &      &     &      &    & &  \\
 Fit-$\hat{\rm II}$  & $6877_{-2}^{+1}$ & $78_{-14}^{+21}$  & $11.1_{-1.1}^{+1.4}$ & $2.0_{-0.7}^{+0.6}$  & $0.06_{-0.01}^{+0.02}$ & $0.01_{-0.00}^{+0.00}$  & $0.07_{-0.01}^{+0.02}$ 
\\ \hline\hline
\end{tabular}
\caption{  Resonance poles of the $X(6900)$ in the third Riemann sheet from the dynamical study including the $\jpsi\jpsi$ and $\psi(3770)\jpsi$ channels are shown. $|\hat{\gamma}_{i=1,2}|$ stand for the resonance couplings to the $\jpsi\jpsi$ and $\psi(3770)\jpsi$ channels, respectively. $\hat{X}_{i=1,2}$ are the partial compositeness coefficients of the $\jpsi\jpsi$ and $\psi(3770)\jpsi$, respectively.    \label{tab.pole2c} } 
\end{table}

\section{Conclusions}

In this work we focus on the narrow peak named $X(6900)$ around 6.9~GeV observed in the $\jpsi\jpsi$ event distributions from the LHCb measurements~\cite{Aaij:2020fnh}, which is the first discovered  fully  heavy-flavored tetraquark candidate. Several different theoretical approaches, including the effective-range expansion, a combination of the compositeness relations, and saturation of the decay width, as well as  unitarized phenomenological amplitudes, are used to investigate the $X(6900)$ state. It is remarkable that different theoretical methods lead to similar conclusions for the $X(6900)$: the $\jpsi\jpsi$, $\xca\xca$, $\xcb\xcb$ and $\psip\jpsi$ components do not play dominant roles in the $X(6900)$. The most important component should be a bare or elementary one, e.g., a compact four-charm-quark state or another microscopic degrees of freedom. This is due to the presence of a CDD pole lying almost on top of the resonance mass, which makes the compositeness $X$ very small and to the applicability of the Morgan's counting-pole criterion with similar poles lying in different Riemann sheets. We have also provided the resulting pole positions, couplings to the different channels and partial as well as total compositeness coefficients. 
A prediction for the coupled-channel dynamics between the $\xca\xca$ and $\xcb\xcb$ channels is the emergence of a virtual state just below the $\xca\xca$ threshold that couples strongly to those channels and weakly to the $\jpsi\jpsi$ channel, making its width is very small. We name this state  $X(6825)$. Its presence could be ascertained experimentally  by improving the energy resolution and the statistics of the $\jpsi\jpsi$ event distribution in future experiments.

Our sophisticated coupled-channel study confirms that two different types of  resonance poles can be obtained for the $X(6900)$, by taking two estimates of the background contributions from the LHCb~\cite{Aaij:2020fnh}. Different line shapes for the $\xca\xca$ and  $\xcb\xcb$ distributions are predicted. Future experimental measurements on the $\xca\xca$ and $\xcb\xcb$ will definitely be helpful to further pinning down the properties of the $X(6900)$.

\section*{Acknowledgements}
We thank Feng-Kun Guo for useful discussions. This work is funded in part by the Natural Science Foundation of China under Grants No.~11975090 and  No.~11575052, the Natural Science Foundation of Hebei Province under Contract No.~A2015205205, and MINECO (Spain) and FEDER (EU) Grants No.  FPA2016-77313-P and MICINN (Spain) No. PID2019-106080GB-C22.

\end{document}